# Charge order and suppression of superconductivity in $HgBa_2CuO_4$ at high pressures

Manuel Izquierdo[1,2], Daniele C. Freitas[3,4], Dorothée Colson[5], Gastón Garbarino[6], Anne Forget[5], Stephan Megtert[7,8], Helène Raffy[7], Robert Comes[8], Jean-Paul Itié[2], Sylvain Ravy[2,8], Pierre Fertey[2] and Manuel Núñez-Regueiro[3]

1  *European XFEL GmbH, Albert Einstein Ring 19, 22761 Hamburg, Germany*

2  *Synchrotron SOLEIL, L'Orme des Merisiers St. Aubin, BP 48, 91192 Gif sur Yvette, France*

3   *Institut Néel, Universit_e Grenoble Alpes and Centre National de la Recherche Scientifique 25 rue des Martyrs - BP 166, 38042, Grenoble cedex 9 France*

4   *Centro Brasileiro de Pesquisas Fisicas, Rua Dr. Xavier Sigaud, 150, Urca, Rio de Janeiro - RJ, Brasil*

5  *CEA Saclay, IRAMIS, SPEC (CNRS URA 2464), F-91191 Gif sur Yvette, France*

6  *European Synchrotron Radiation Facility (ESRF), BP 220, 38043 Grenoble Cedex 9, France* 7   *Unité Mixte de Physique, UMR137 CNRS/Thal_es, F-91767 Palaiseau Cedex, France*

8  *Laboratoire de Physique des Solides, Université Paris-Sud 11, CNRS UMR 8502, 91405 Orsay Cedex, France*

New insight into the superconducting properties of $HgBa_2CuO_4$ (Hg-1201) cuprates is provided by combined measurements of the electrical resistivity and single crystal X-ray diffraction under pressure. The changes induced by increasing pressure up to 20GPa in optimally doped single crystals were investigated. The resistivity measurements as a function of temperature show a metallic behavior up to ~10GPa that gradually passes to an insulating state, typical of charge ordering, that totally suppresses superconductivity above 13GPa. The changes in resistivity are accompanied by the apparition of sharp Bragg peaks in the X-ray diffraction patterns indicating that the charge ordering is accompanied by a 3D oxygen ordering appearing at 10GPa of wavevector [0.25, 0, L]. As pressure induces a charge transfer of about 0.02 at 10GPa, our results are the first observation of charge order competing with superconductivity that develops in the over-doped region of the phase diagram of a cuprate.

In recent years, the observation of charge order[1-3] in cuprates other than the well-studied one[4] of $La_{2-x}Ba_xCuO_4$, has lead to a boom of studies in the hope of finding the key in the understanding of high temperature superconductors. Of particular interest are the charge density wave (CDW) fluctuations observed in *Hg-1201*[5,6]. Contrary to other families, the mercury family has flat tetragonal $CuO_2$ planes, and there is apparently no plane distortion but an oxygen ordering that determines the charge density wave. Most interestingly, the [H] component of the CDW wave vector has been shown to scale[6] with those determined for the YBCO system as a function of doping. Diffraction studies have shown that the oxygen ordering has one dimensional character that manifests as diffuse lines due to fluctuating charge ordering in the two tetragonal directions[7,8]. More recently, a phase separation scenario has been proposed by scanning micro X-ray diffraction in which CDW regions and oxygen interstitial regions coexist[9].

In order to understand the relevance of oxygen ordering in the superconductivity of *Hg-1201*, which exhibits only slight intensity modifications upon changing temperature[7] we decided to perform experiments under pressure. This variable has been shown to increase the transition temperature in cuprates but particularly in the case of mercury compounds. Thus, the highest superconducting transition temperatures ($T_c$) reported so far, $T_c$ =166K, was on *Hg1223-F* at 25GPa[10]. The mechanisms controlling this increase have been thoroughly studied[11-14]. Normally, the leading mechanism is the charge transfer under pressure, due to the strong compression along the *c* axis, that reduces the ionicity of the layers and causes a passage of electrons from the negatively charge $CuO_2$ layers to the reservoir layers. The result is a parabolic variation of $T_c$ under pressure. The compression of the *a* parameter involving a significant shortening of the *CuO* bond[15], can induce a strong linear variation of $T_c$, that has been used to explain the anomalously strong increase observed in the flat $CuO_2$ plane *Hg* cuprates[16]. It should also affect the fluctuating one dimensional oxygen ordering recently reported. Furthermore, a correlation between $T_c$ and the changes in oxygen ordering should show up when the latter is relevant for superconductivity.

We have therefore performed electrical resistivity and single crystal X-ray diffraction studies under pressure on an optimally doped *Hg-1201*. Single crystals were synthesized using a flux technique by identifying the most favorable region of the ternary diagram *HgO-BaO-CuO* to get *Hg-1201* single crystals. They have well-developed (001) faces with very clean surfaces and a size in the range of 0.3x 0.3x 0.3 mm$^3$. The critical temperature, measured with a SQUID magnetometer, showed a transition onset at $T_c$ =95K and a narrow width (~4K) for isolated single crystals, thus confirming a high sample quality[18].

Electrical resistivity was measured in a solid state pressure cell. In Fig. 1(b) the electrical resistance of a *Hg-1201* single crystal as a function of temperature and pressure is displayed. At the lowest pressure, the behavior is clearly metallic with a sharp superconducting transition. As pressure increases, the electrical resistivity decreases up to about 5GPa. At higher pressures the resistivity starts increasing and the superconducting transition widens. The last faint signature of a superconducting transition is observable at 11.5GPa. At higher pressures, the sample shows an activated behavior typical of an insulator. This can be due either to some sort of pressure induced ordering or sample degradation. Even though the solid-state pressure cell is not conceived to measure on decompression, we have performed a measurement at 4GPa on decompression. We observe that the sample has recovered the superconducting state, but not the metallic character. This can be due to sample degradation at high pressures, to the non-homogenous strains due to the decompression of the solid-state cell or both Fig. 1(c).

In Fig. 1(d) we show the evolution of the resistivity with pressure. Up to 5GPa $T_c$ increases with a slope of 1.2K/GPa, a value slightly lower than previously reported for nominally optimal doped samples[20,21]. This indicates that the investigated samples are probably well on the summit of the doping parabola. Above 5GPa $T_c$ starts decreasing monotonically reaching a zero value at 15GPa. The $T_c$ on decompression is also shown, and almost coincides with the one obtained upon compression. We could not determine a transition temperature towards an ordering that would explain the activated behavior of the resistance, probably due to pressure inhomogeneities, as is often the case in this type of cells. However, we can quantify the passage to an insulating state by plotting the ratio of the low temperature resistance to the ambient temperature resistance. We plot this evolution and we are able to plot it with a power law mean field expression $[1 - P/P_c]^{0.5}$, with $P_c$=11GPa.

In Fig. 2 we show diffraction patterns on another *Hg-1201* measured in a diamond anvil cell at 24.5 KeV at the CRISTAL beamline of synchrotron Soleil.

Three different type of diffraction patterns as a function of pressure can be distinguished. Below 7.5GPa we see well defined tetragonal spot that increase in intensity with pressure. They show the very good crystal quality of the crystal measured. Furthermore, the diffuse streaks, already described as corresponding to fluctuating 2D linear oxygen chains, are also visible[7]. The intensity of the streaks increases around the (±1 ±1 L). In the range from 8GPa to 13GPa new spots appear only at well defined positions along the tetragonal directions in reciprocal space. From there on, the number of spots increase but some of them do not lie anymore along the initial tetragonal directions. Furthermore, the tetragonal spots become much weaker and elongated indicating a strong modification of structure of the sample. Strong modifications are observed along the (0 K L) and (H 0 L) where the initial diffuse lines seem to develop into intense, wide and well defined spots. Their incommensurate periodicity seems to be unrelated to that of the extra spots at lower pressure values thus supporting the phase separation scenario.

One can understand the apparition of incommensurable spots by the increase the correlation between the oxygen chains responsible of the diffuse lines. If this is indeed the case, and since no extra diffuse lines develop in the *a-b* plane, the ordering has to take place along the *c*-axis. The indexation of the peaks in the high pressure range shows that the periodicity corresponds to 8 unit cells along the *c* axis. Correlations along the *c*-axis have been previously reported at low pressure on Hg-1223 cuprates. However, in this case the reported superstructure had a 5*c* periodicity[17]. Since in our diffraction patterns incommensurable spots only appear along particular segments of the diffuse lines this leaves room to another interpretations. One possibility would be the formation of orthorhombic twin domains upon applying pressure. This type of domains has been observed in YBCO compounds giving rise to diffraction spots at variable distance from the tetragonal spots. Another possibility, deriving for the phase separation scenario recently proposed by scanning micro x-ray diffraction studies would be that the incommensurate spots have two origins. Part of them would be related to pressure induced ordering of the oxygen atoms and the other part to that of the CDW regions.

We have quantified the evolution of the superstructure order by plotting the intensity of the new peaks normalized to the intensity of the neighboring tetragonal Bragg peaks. The results displayed on the left panel of Fig. 3 show their evolution with pressure compared with that of $T_c$. This proves that the development of the 3D oxygen ordering destroys the superconducting state by generating a charge order that explains the insulating state observed at high pressure in our resistance measurements.

On the other hand, we can estimate[10] the additional doping introduced application of 10GPa to be between around 0.02, as dn/dP 0.002h/GPa. As we started with an optimally doped mono crystal, at p=0.18 we are clearly in the overdoped region. Thus, in strong contrast to all other previous reported charge ordering in cuprates competing strongly superconductivity, only observed in the pseudogap underdoped region, we observed it in the overdoped Fermi-liquid region.

Our measurements pose the following fundamental question: as it is observed in a region where the compounds should be in a "normal" Fermi-liquid state, has charge ordering something to do with the mechanism of high temperature superconductivity or is it just a phenomenon related to the layered structure of cuprates that has nothing to do with the mechanism of superconductivity? In particular, it was proposed long ago that the low temperature phases of the $La_{2-x}Ba_xCuO_4$ system are the result of different thermal contractions of the $CuO_2$ and $LaO$ layers[19]. This mechanism would be unrelated to the mechanism of high temperature superconductivity, and can appear at different concentrations. In this line of thought, one can even wonder if the mysterious pseudogap region is just the result of the coupling of the interaction responsible for the high $T_c$ with the structural phenomenon and not an intrinsic property for high temperature superconductivity. On the other hand, it can be part of the complexity necessary for the appearance of high temperature superconductivity[22]. On the other hand, it is known that a coherence transition temperature $T_{coh}$ is expected in the overdoped region, with a power law dependence starting from a putative quantum critical point (QCP) whose exact position on the concentration axis is still a matter of controverse. Our observations could be in a symmetric region of the p = 0.125 magical number with respect to the QCP and future calculations might explain our results. Furthermore, the relation of our observations with the CDWs observed at lower temperatures[6,9] remains to be established. In this respect, pressure dependent experiments as a function of the temperature are envisaged.

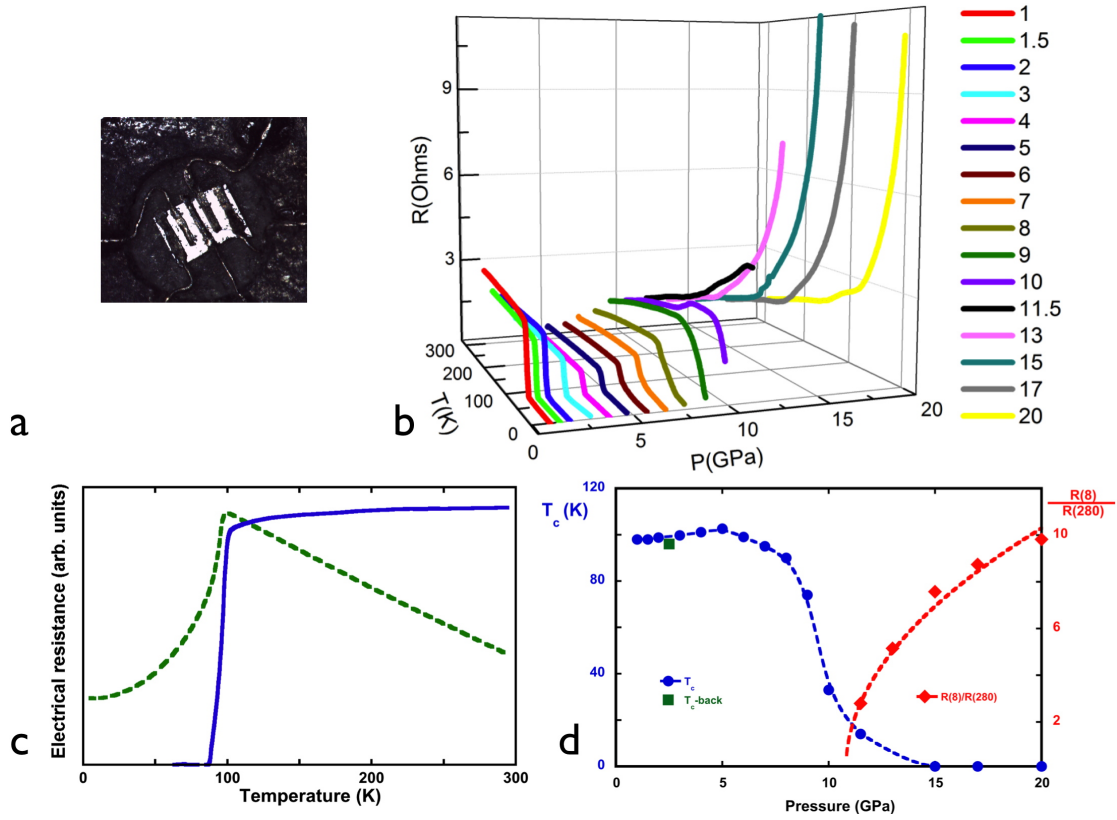

**Figure 1.**

(a) Single crystal of *Hg-1201* mounted in the pressure cell. (b) Electrical resistance of the *Hg-1201* single crystal as a function of pressure and temperature. The evolution from a metallic and superconducting behavior to an insulating (indicating charge-order) and non-superconducting one at high pressures is evident. (c) Comparison of the resistivities at 4GPa on compression and decompression. The superconducting behavior is recovered although the decompressed sampledoes not show a clear metallic behavior. (d) In blue circles we show the evolution of $T_c$ with pressure. It increases up to about 5GPa and further decreases attaining zero at 15GPa. The red diamonds correspond to the ratio R(8K/R(280), a way ofshowing the increase of the charge ordering with pressure. The dashed red curve is a mean-field power law fit.

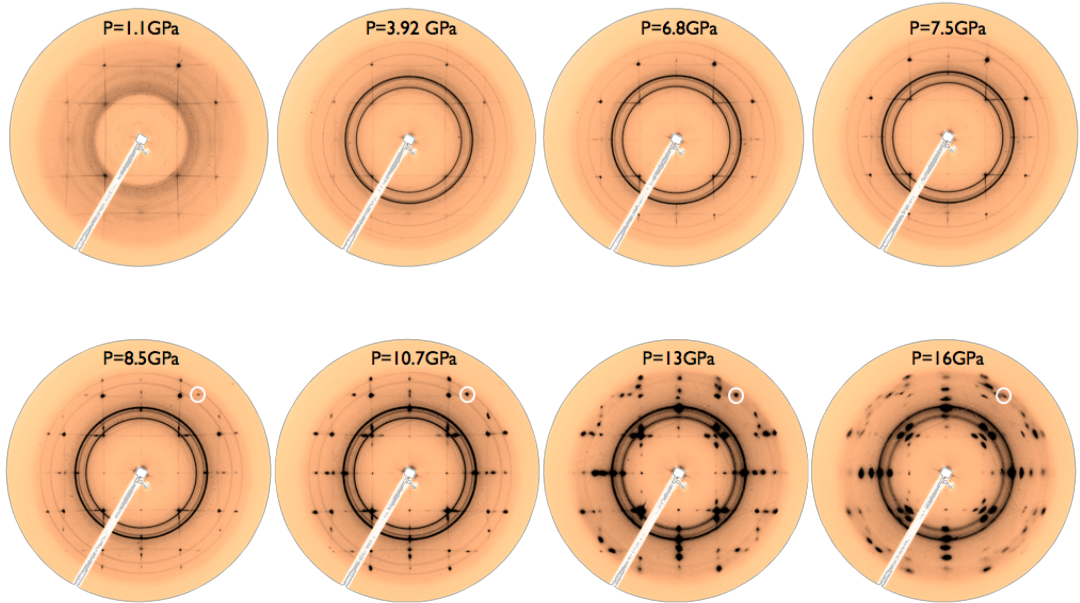

**Figure 2.**

Diffraction patterns as a function of pressure. We observe a well defined tetragonal structure at low pressures, with diffuse stripes that indicate a fluctuating 2D linear oxygen ordering[7]. Above 8GPa we observe the appearance of superstructure spots (marked by white circles). They become more intense and diffuse at very high pressures.

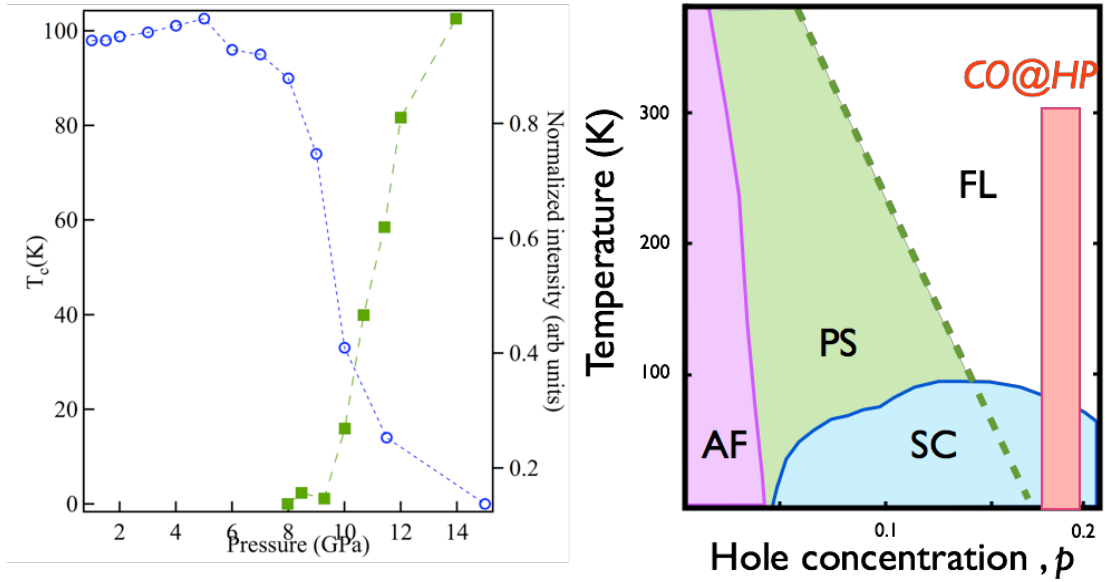

**Figure 3.**

(left panel) Normalized intensity of the superstructure spots as a function of pressure compared to the evolution of $T_c$. (right panel) Phase diagram of mercury cuprates showing where our results are situated as a function of doping.